\documentclass{llncs}
\usepackage{url}
\usepackage{hyperref}
 \hypersetup{
 	colorlinks,
 	citecolor=red,
 	filecolor=blue,
 	linkcolor=darkgray,
 	urlcolor=red
 }

\usepackage{graphicx} 
\usepackage[%
    font={small,sf},
    labelfont=bf,
    format=hang,    
    format=plain,
    margin=0pt,
    width=\textwidth,
]{caption}
\usepackage{subcaption}
\usepackage[list=true]{subcaption}
\usepackage{paralist}
\usepackage{color}
\usepackage{xcolor}
\usepackage{wrapfig } 
\usepackage{booktabs}
\usepackage{multido}
\usepackage{multirow}
\usepackage{bigstrut}
\usepackage{longtable}
\usepackage[inline]{enumitem} 
\usepackage{rotating}
\usepackage{lipsum} 

\usepackage{graphicx}
\usepackage[colorinlistoftodos]{todonotes}
\usepackage{xspace} 
\usepackage{framed} 
\usepackage{mdframed}

\newcommand{\openmp}{Open{MP}\xspace}

\newcommand{\tss}{TSS\xspace}

\newcommand{\facTWO}{FAC2\xspace}
\newcommand{\wfTWO}{WF2\xspace}
\newcommand{\random}{RAND\xspace}
\newcommand{\static}{STATIC\xspace}
\newcommand{\gss}{GSS\xspace}
\newcommand{\selfsched}{SS\xspace}

\newcommand{\longtss}{trapezoid self-scheduling\xspace}
\newcommand{\longfacTWO}{factoring\xspace}
\newcommand{\longwfTWO}{weight\-ed factoring\xspace}
\newcommand{\longrandom}{random\xspace}

\newcommand{\ac}{\texttt{ac}\xspace}
\newcommand{\cmd}{\texttt{c\_md}\xspace}
\newcommand{\lava}{\texttt{lava.md}\xspace}
\newcommand{\tfomd}{\texttt{350.md}\xspace}
\newcommand{\nasmg}{\texttt{NAS MG}\xspace}
\newcommand{\cut}[1]{\xspace}

\newcommand{\secref}[1]{\ref{#1}\xspace}
\newcommand{\figref}[1]{Fig.\,\ref{#1}\xspace}
\newcommand{\Figref}[1]{Fig.\,\ref{#1}\xspace}

\newcommand{\parahint}[1]{}

\newcommand{\oursubsection}[1]{\par\textbf{#1:}\xspace}
\begin{document}

\pagestyle{plain}
%
%
\title{OpenMP Loop Scheduling Revisited: \\Making a Case for More Schedules}
\titlerunning{DLS in OpenMP}  
%
\author{Florina~M.~Ciorba\inst{1} \and
Christian Iwainsky\inst{2}\and
Patrick Buder\inst{1}\textsubscript{\footnotemark[4]}\footnotetext[4]{Authors in order of contribution.}}

\authorrunning{Florina M. Ciorba et al.} 
%
\tocauthor{Florina M. Ciorba, Christian Iwainsky, and Patrick Buder$*$}


%
\institute{
University of Basel, Switzerland\\
\email{\{florina.ciorba,p.buder\}@unibas.ch}\\ 
\texttt{hpc.dmi.unibas.ch}
\and
Technische Universit\"at Darmstadt, Germany\\
\email{christian.iwainsky@sc.tu-darmstadt.de}\\ 
\texttt{www.sc.tu-darmstadt.de | www.hkhlr.de}
}
\setcounter{footnote}{0}
\maketitle              
\setcounter{footnote}{0}
\

\begin{abstract}
In light of continued advances in loop scheduling, this work revisits the \openmp loop scheduling by outlining the current state of the art in loop scheduling and presenting evidence that the existing \openmp schedules are insufficient for all combinations of applications, systems, and their characteristics. 
A review of the state of the art shows that due to the specifics of the parallel applications, the variety of computing platforms, and the numerous performance degradation factors, no single loop scheduling technique can be a `one-fits-all' solution to effectively optimize the performance of all parallel applications in all situations.
The impact of irregularity in computational workloads and hardware systems, including operating system noise, on the performance of parallel applications results in performance loss and has often been neglected in loop scheduling research, in particular the context of \openmp schedules.
Existing \emph{dynamic loop self-scheduling techniques}, such as \longtss, \longfacTWO and \longwfTWO, offer an unexplored potential to alleviate this degradation in \openmp due to the fact that they explicitly target the minimization of load imbalance and scheduling overhead.
Through theoretical and experimental evaluation, this work shows that these loop self-scheduling methods provide a benefit in the context of \openmp.
In conclusion, \openmp must include more schedules to offer a broader performance coverage of applications executing on an increasing variety of heterogeneous shared memory computing platforms.
\end{abstract}
\section{Introduction}
Loop-level parallelism is a very important part of many OpenMP programs. 
OpenMP\,\cite{OpenMP} is the decades-old industry standard for parallel programming on shared memory platforms. 
Due to wide support from industry and academia, a broad variety of applications from science, engineering, and industry are parallelized and programmed using OpenMP\,\cite{OpenMP-API:2017}. 

Applications in science, engineering, and industry are complex, large, and often exhibit irregular and non-deterministic behavior. 
Moreover, they are frequently \emph{computationally-intensive} and consist of large \emph{data parallel loops}. 
High performance computing platforms are increasingly complex, large, heterogeneous, and exhibit massive and diverse parallelism. 
The optimal execution of parallel applications on parallel computing platforms is \mbox{NP-hard}\,\cite{NP-Hard:1990}.
This is mainly due to the fact that the individual processing times of application tasks\footnote{\emph{Tasks} and \emph{loop iterations} denote independent units of computation and are used interchangeably in this work.}\xspace cannot, in general, be predicted, in particular on machines with complex memory hierarchies (e.g., non-uniform memory access)\,\cite{FSC-DataLoc:1996}. 

The performance of applications can be degraded due to various ``overheads'', which are synchronization, management of parallelism, communication, and load imbalance\,\cite{Banicescu:1996}.
Indeed, these overheads cannot be ignored by any effort to improve the performance of applications, such as the loop scheduling schemes\,\cite{BAL:1998}.

Load imbalance is the major performance degradation overhead in com\-pu\-ta\-tion\-ally-intensive applications\,\cite{FRAC:1996,IESP-2.0}. 
It can result from the uneven assignment of computation to units of work (e.g., threads) or the uneven assignment of units of work to processors. 
The former can be mitigated via a fine-grained decomposition of computation into units of work. 
The latter is typically minimized via the use of a central (ready) work queue from which idle processors remove units of work. 
This approach is called \emph{self-scheduling}.
The use of a central work queue facilitates a dynamic and even distribution of load among the processors and ensures that no processor remains idle while there is work do be conducted. 
The dynamic nature of the self-scheduling schedules combined with their centralized work queue characteristic makes them \emph{ideal} for use in \openmp.
Self-scheduling is already supported in \openmp by the scheduling mechanisms of parallelizing work constructs, namely the \texttt{parallel fo}r loop constructs.
The scheduling responsibility falls onto the \openmp runtime system, rather than on the operating system or the (potentially scheduling non-expert) programmer.
The \openmp runtime system can be precisely optimized for scheduling and shared memory programming, while the operating system must be generic to accommodate a variety of programming models and languages.   

No single loop scheduling technique can address all sources of \emph{load imbalance} to effectively optimize the performance of all \emph{parallel applications} executing on all types of \emph{computing platforms}. 
Indeed, the characteristics of the loop iterations \emph{compounded} with the characteristics of the underlying computing systems determine, typically during execution, whether a certain scheduling scheme outperforms another. 
The impact of system-induced variability (e.g., operating system noise, power capping, and others) on the performance of parallel applications results in additional irregularity and has often been neglected in loop scheduling research, particularly in the context of OpenMP schedules\,\cite{DLS+API:2003,Automatic-OMP-LS:2012}. 

There exists a great body of work on loop scheduling and a taxonomy of scheduling methods is included in Section~\ref{sec:revisiting}.
In essence, the present work makes the case that the existing OpenMP schedules (\texttt{static}, \texttt{dynamic}, and \texttt{guided}) are insufficient to cover all combinations of applications, systems, and variability in their characteristics\,\cite{DLS+API:2003,Adaptive-OMP-LS:2004}. 

The present work revisits the loop schedules in the OpenMP specification\,\cite{OpenMP}, namely \texttt{static}, \texttt{dynamic}, and \texttt{guided}\,\cite{GSS:1987}, and challenges the assumption that these are sufficient to efficiently schedule all types of applications with parallel loops on all types of shared memory computing systems.
Moreover, this work makes the case that \openmp must include more schedules, and proposes the \emph{loop self-scheduling} class, to offer a broader performance coverage of OpenMP applications executing on an increasing variety of shared memory computing platforms. 
The additional loop scheduling techniques considered herein are trapezoid self-scheduling\,\cite{TSS:1993}, factoring\,\cite{FAC:1991}, weighted factoring\,\cite{WF:1996}, and random. 
The four dynamic loop self-scheduling (DLS) techniques have been implemented in the LaPeSD libGOMP\,\cite{LaPeSD-LibGOMP} based on the GNU \openmp library and evaluated using well-known benchmarks.
The experimental results indicate the feasibility of adding the four loop \mbox{self-scheduling} techniques to the existing \openmp schedules.
The results also indicate that the existing \openmp schedules only partially covered the achievable performance spectrum for the benchmarks, system, and pinnings considered. 
The experiments confirm the original hypothesis and that certain DLS outperform others without a single DLS outperforming all others in all cases. 
Thus, the results strongly support the case of ample available room for improving applications performance using more \openmp schedules.

The remainder of this work is organized as follows. 
Section~\ref{sec:related-work} reviews the work related to implementing various loop schedules into compilers and runtime systems. 
Loop scheduling is revisited in Section~\ref{sec:revisiting}, with a focus on the class of dynamic loop self-scheduling techniques and the existing OpenMP schedules. 
Section~\ref{sec:dls-in-omp} describes the implementation of the dynamic loop self-scheduling techniques considered in this work, the selection of the benchmarks used to evaluate their performance, as well as offering details of the measurement setup and the experiments.
This work concludes in Section~\ref{sec:conclusion} with a discussion and highlights of future work aspects. 

\section{Related Work}\label{sec:related-work}
The motivation behind the present work is reinforced by the plethora and diversity of the existing related work, indicating that scheduling of loops in \openmp is a long-standing and active area of research with diverse opportunities for improvement. 
The most recent and relevant efforts of implementing additional scheduling methods into parallelizing compilers and runtime systems are briefly discussed next.

A preliminary version of this work\,\cite{LIBGOMP+DLS:2017} contains the first prototype implementation, in LaPeSD libGOMP\,\cite{LaPeSD-LibGOMP}, of three self-scheduling techniques considered herein, trapezoid self-scheduling\,\cite{TSS:1993}, factoring\,\cite{FAC:1991}, and weighted factoring\,\cite{WF:1996}, while random is a newly added schedule in the present work.
The major focus therein was the feasibility of implementing new loop self-scheduling techniques in libGOMP.
An exploratory evaluation of the self-scheduling techniques for a broad range of \openmp benchmarks from various suites confirmed that static scheduling is beneficial only for loops with uniformly distributed iterations; it also confirmed the need for dynamic scheduling of loops with irregularly distributed iterations.
Most of the loop schedules proposed over time in \openmp employ affinity-based principles\,\cite{SAS+AFS+FGDLS+OpenMP:2003}, while the remaining employ work stealing\,\cite{AdaptiveOMP-LS:2013,UDefSched:2017} or other variations of static scheduling\,\cite{UDefSched:2017,BinLPT:2017,SRR:2017}.
A number of approaches rely on profiling information about the application on a given architecture\,\cite{JIT-LB:2013,Automatic-OMP-LS:2012,RES-DLS:2005}. 
Only very little work has considered additional loop self-scheduling methods\,\cite{DLS+API:2003,Adaptive-OMP-LS:2004}, while the present work is the first to consider weighted factoring\,\cite{WF:1996} and random self-scheduling.
The impact of system-induced variations has only been considered in limited instances\,\cite{Automatic-OMP-LS:2012}, whereas the present work considers the \emph{cumulative impact} of variabilities in problem, algorithm, and system. 
Most related work considers benchmarks from well-known suites.
The present work provides targeted experiments for a particular class of real-world applications, namely molecular dynamics.
Existing efforts also implemented their scheduling prototypes in \mbox{libGOMP}\,\cite{Automatic-OMP-LS:2012,JIT-LB:2013,SRR:2017,BinLPT:2017,LIBGOMP+DLS:2017}, similar to this work.
No studies explore various pinning strategies, which are used in this work to highlight the benefit of dynamic loop self-scheduling.
Moreover, in this work experiments are conducted with 20 threads on a state-of-the-art 10$\times$2-way cores Intel Broadwell architecture. 
\section{Revisiting Loop Scheduling}\label{sec:revisiting}

\begin{figure}[!h]
\centering
\subcaptionbox{Load balancing via dynamic loop \mbox{self-scheduling}\label{fig:DLB}}
{\includegraphics[width=0.5\textwidth]{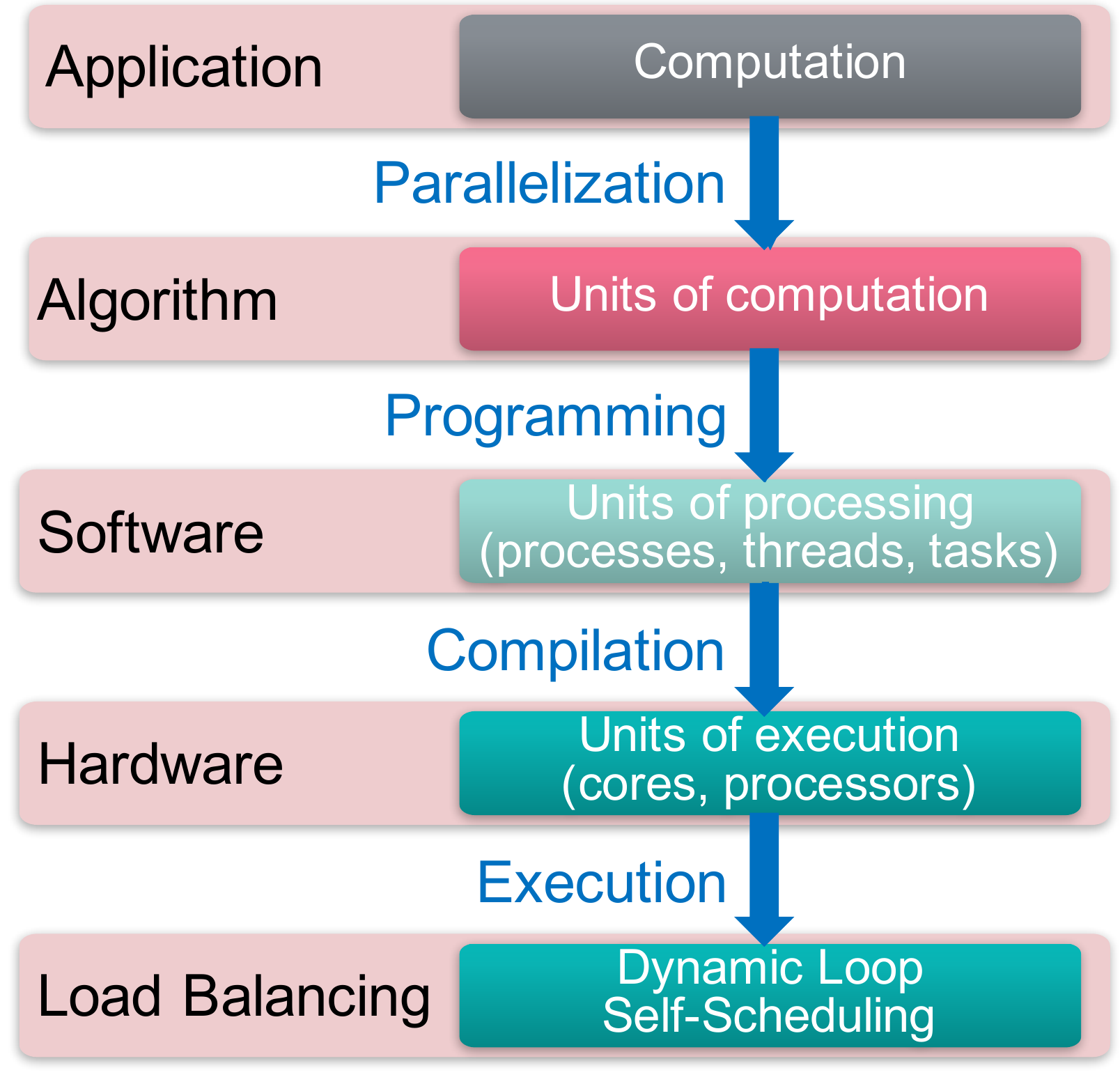}}%
\vspace{0.5cm}
\hfill%
\subcaptionbox{Taxonomy of loop scheduling by assignment, scheduling decisions, and target optimization goals\label{fig:taxonomy}}
{\includegraphics[width=0.95\textwidth]{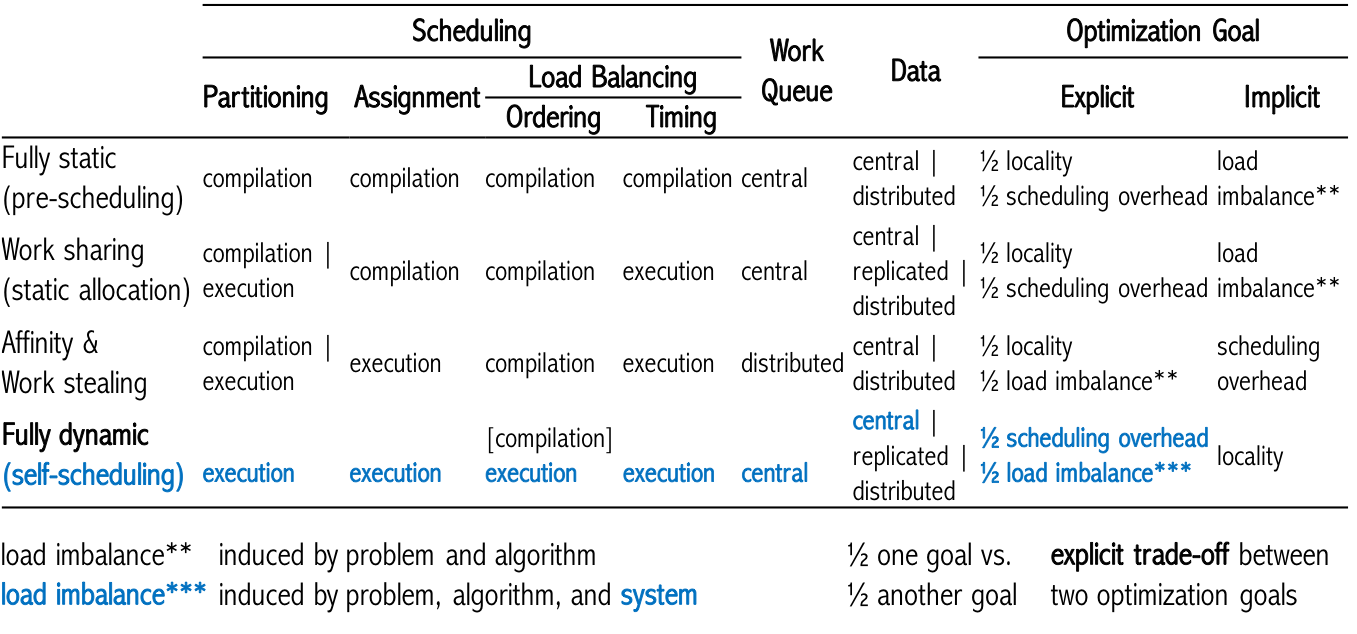}}%
\caption{Loop scheduling and its relation to load balancing, overhead, and locality.}
\end{figure}

\emph{Parallelization} is the process of identifying units of computations that can be performed in parallel on their associated data, as well as the ordering of computations and data, in space and time, with the goal of decreasing the execution time of applications.
The process of parallelization consists of three steps: partitioning (or decomposition), assignment (or allocation), and scheduling (or ordering). 
\emph{Partitioning} refers to the decomposition of the computational domain and of the corresponding computations and data into parallel subparts, called \emph{units of computation} (see Fig.~\ref{fig:DLB}). 
Through programming, the units of computation are mapped to software \emph{units of processing}, such as processes, threads, or tasks\footnote{Task denotes, in this context, the form of concurrency at the level of a programming paradigm, e.g., OpenMP 3.0 tasks.} (see Fig.~\ref{fig:DLB}). 
\emph{Assignment} refers to the allocation of the units of processing to \emph{units of execution} in space, such as cores or processors. 
The allocation can be performed \emph{statically}, during compilation or before the start of the application execution, or \emph{dynamically}, during application execution (see Fig.~\ref{fig:taxonomy}).  
\emph{Scheduling} refers to the ordering and timing of the units of computations to achieve a load balanced execution. 
The ordering can be \emph{static}, determined at compilation time, \emph{dynamic}, performed during application execution, or \emph{adaptive}, which changes either the static or the dynamic ordering during execution (see Fig.~\ref{fig:taxonomy}).  

\emph{Load imbalance} is the major performance degradation factor in com\-pu\-ta\-tion\-ally-intensive applications\ \cite{IESP-2.0}.
Applications suffer from load imbalance when certain processors are idle and yet there is work ready to be performed that no processor has started. 
Load imbalance can result from the uneven assignment of computation to units of work (e.g., threads) or the uneven assignment of units of work to processors (see Fig.~\ref{fig:DLB}). 
The causes of both types of uneven assignment can be due to the \emph{problem} (e.g., non-uniform data distribution), \emph{algorithm} (boundary phenomena, convergence, conditions and branches), or induced by the \emph{system} (memory access interference,  operating system noise, or use of shared resources). 
The uneven assignment of computation to units of work can be mitigated via a fine-grained decomposition of computation into units of work. 
The uneven assignment of units of work to processors is typically minimized via the use of a central (ready) work queue from which idle and available processors remove units of work. 
This approach is called \emph{self-scheduling}.
Self-scheduling facilitates a dynamic and even distribution of load among the processors and ensures that \emph{no processor remains idle} while there is work to be conducted. 

\emph{Loop scheduling} refers to the third parallelization step, described above, applied to the iterations of a loop. 
As illustrated in Fig.\,\ref{fig:taxonomy}, depending on the assignment, the ordering and timing of the scheduling decisions, as well as the target optimization goals, existing loop scheduling solutions can be classified into: 
fully static or \emph{pre-scheduling}, 
\emph{work sharing}, 
\emph{affinity} and \emph{work stealing}, 
and fully dynamic or \emph{self-scheduling}. 
In the \emph{absence} of load imbalance, approaches that employ pre-scheduling or static allocation are highly effective as they favor data locality and have virtually no scheduling overhead\,\cite{LDS:1993}.
Affinity-based scheduling strategies\,\cite{SPAFS:1994,DAFS-WPAFS:1994,AdaptiveOMP-LS:2013,UserDefinedSched:2015} also deliver high data locality. 
In the \emph{presence} of load imbalance, which is the major application performance degradation factor in computationally-intensive applications, strategies based on work stealing only mitigate load imbalance caused by problem and algorithmic characteristics\,\cite{WorkStealing:1999,AdaptiveOMP-LS:2013}.
Both affinity and work stealing incur non-negligible scheduling overhead. 
Scheduling that employs affinity and work stealing only partially addresses load imbalance and trades it off with data locality. 
None of the aforementioned scheduling approaches explicitly minimize load imbalance \emph{and} scheduling overhead.
\emph{Fully dynamic self-scheduling} techniques, such as those considered in this work, explicitly address \emph{all} sources of load imbalance (caused by problem, algorithmic, and systemic characteristics) and attempt to minimize the scheduling overhead, while implicitly addressing allocation delays and data locality\,\cite{AF:2000,BAL:2002,KASS:2012}. 

This work considers dynamic loop \emph{self-scheduling} in the context of OpenMP scheduling, to explicitly address \emph{load imbalance} and \emph{scheduling overhead} for the purpose of minimizing their impact on application performance.
The OpenMP specification\,\cite{OpenMP} provides three types of loop schedules: \texttt{static}, \texttt{dynamic}, and \texttt{guided}, which can be directly selected as arguments to the OpenMP \texttt{parallel for schedule()} clause.
The loop schedules can also automatically be selected by the OpenMP runtime via the \texttt{auto} argument to \texttt{schedule()} or their selection can be deferred to execution time via the \texttt{runtime} argument to \texttt{schedule()}.

The use of \texttt{schedule(static,chunk)} employs \emph{straightforward parallelization} or \emph{static block scheduling}\,\cite{LDS:1993} (STATIC) wherein $N$ loop iterations are divided into $P$ chunks of size $\lceil N/ P \rceil$; $P$ being the number of processing units. 
Each \texttt{chunk} of consecutive iterations is assigned to a processor, in a round-robin fashion.
This is only suitable for \emph{uniformly distributed} loop iterations and in the \emph{absence} of load imbalance.
The use of \texttt{schedule(static,1)} implements \emph{static cyclic scheduling}\,\cite{LDS:1993} wherein single iterations are statically assigned consecutively to different processors in a cyclic fashion, i.e., iteration $i$ is assigned to processor $i$ \texttt{mod} $P$. 
For certain non-uniformly distributed parallel loop iterations, cyclic produces a more balanced schedule than block scheduling. 
Both versions achieve high locality with virtually no scheduling overhead, at the expense of poor load balancing if applied to loops with irregular loop iterations or in systems with high variability. 
The dynamic version of \texttt{schedule(static,}\texttt{chunk)} which employs \emph{dynamic block scheduling} is \texttt{schedule(dynamic,chunk)}.
The only difference is that the assignment of chunks to processors is performed during execution.
The dynamic version of \texttt{schedule(static,1)} is \texttt{schedule(dynamic,1)} which employs \emph{pure} \emph{\mbox{self-scheduling}} (PSS), the easiest and most straightforward dynamic loop self-scheduling algorithm\,\cite{PSS:1986}. 
Whenever a processor is idle, it retrieves an iteration from a central work queue.
PSS\footnote{For simplicity and consistency with prior work, PSS is herein denoted SS.} achieves good load balancing yet may introduce excessive scheduling overhead.
\emph{Guided self-scheduling}(GSS) \,\cite{GSS:1987} is implemented by \texttt{schedule(guided)}\-, one of the early self-scheduling techniques that trades off load imbalance and scheduling overhead.

The above OpenMP loop schedules are insufficient to cover the needs for efficient scheduling of all types of applications with parallel loops on all types of shared memory computing systems. 
This is mainly due to overheads, such as synchronization, management of parallelism, communication, and load imbalance, that cannot be ignored\,\cite{BAL:1998}.

This work proposes the use of self-scheduling methods in OpenMP to offer a broader performance coverage of OpenMP applications executing on an increasing variety of shared memory computing platforms.

Specifically, the loop self-scheduling techniques to consider are \emph{trapezoid self-scheduling} (TSS)\,\cite{TSS:1993}, \emph{factoring ``2''} (FAC2)\,\cite{FAC:1991}, \emph{weighted factoring ``2''} (WF2)\,\cite{WF:1996}, and \emph{random}. 
It is important to note that the \facTWO and \wfTWO methods evolved from the probabilistic analysis that gave birth to FAC and WF, respectively, while \tss is a deterministic self-scheduling method.
Moreover, \wfTWO can employ workload balancing information specified by the user, such as the capabilities of a heterogeneous hardware configuration. %

\begin{figure}[!h]
\centering
\includegraphics[width=0.65\textwidth]{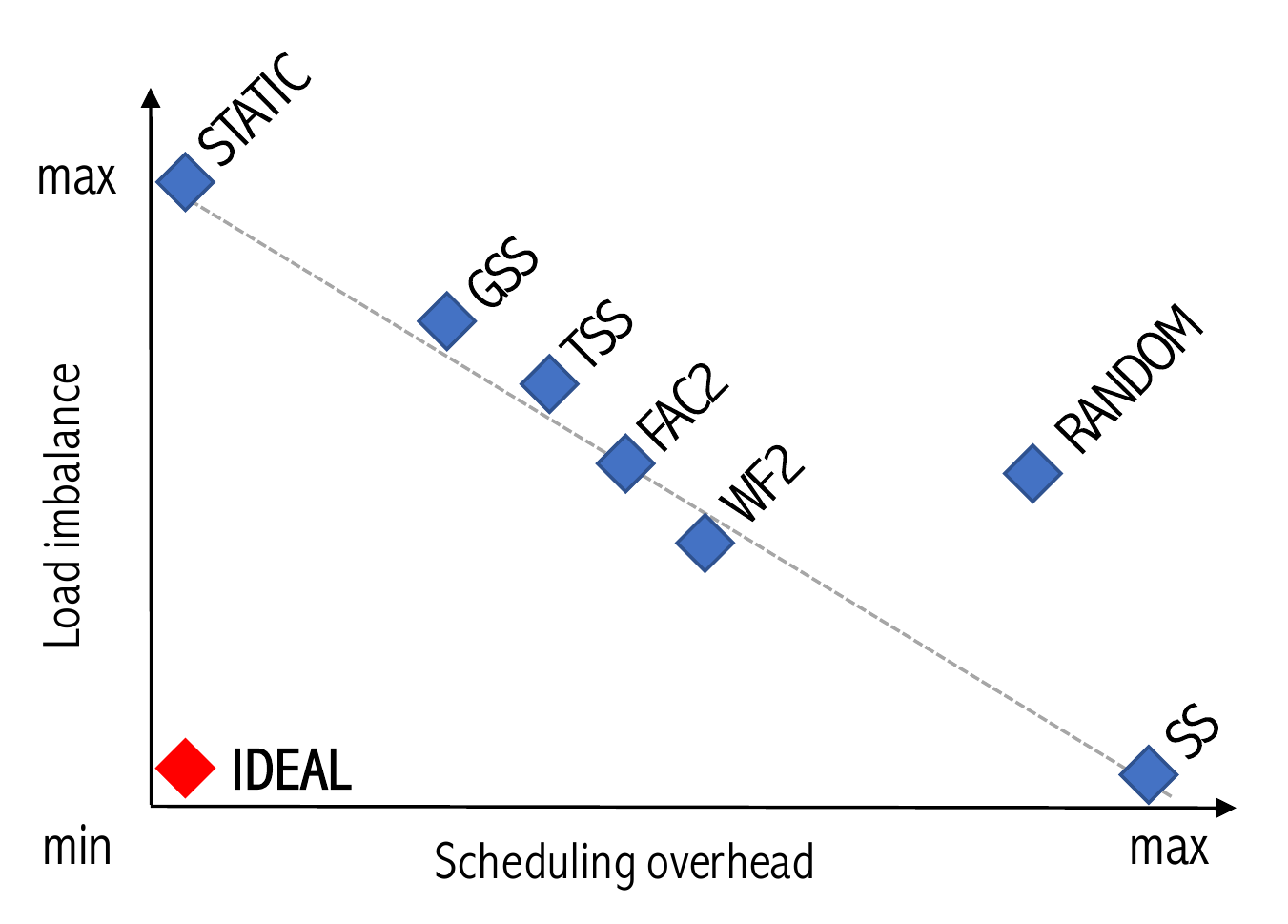}
\caption{\label{fig:ss-extremes}Load imbalance vs. scheduling overhead trade-off for various dynamic loop \mbox{self-scheduling} schemes.}
\end{figure}

Based on their underlying models and assumptions, these self-scheduling techniques are expected to trade load imbalance and scheduling overhead as illustrated in Fig.~\ref{fig:ss-extremes}.

\emph{Random} is a self-scheduling-based method that employs the uniform distribution between a lower and an upper bound to arrive at a randomly calculated chunk size between these bounds. 
The lower bound is equal to the ratio between the total number of iterations and 100 $\times$ the total number of cores.
The upper bound is equal to the total number of iterations divided by 2 $\times$ the total number of cores. 
The quantities 100 and 2 are chosen arbitrarily to denote the fact that the randomly generated chunk size is at least \textit{100 times} and at most \textit{half} the remaining iterations, respectively. 
The smallest randomly generated chunk size, denoted $K_{min}$ is bounded below by 1, i.e., $K_{min} \geq 1$, while the largest randomly generated chunk size, denoted $K_{max}$ is bounded below as $K_{min} \geq K_{min} + 1$.

\begin{figure}[t]
\centering
\subcaptionbox{Chunk sizes\label{fig:DLS-chunks}}
{\includegraphics[width=0.24\textwidth]{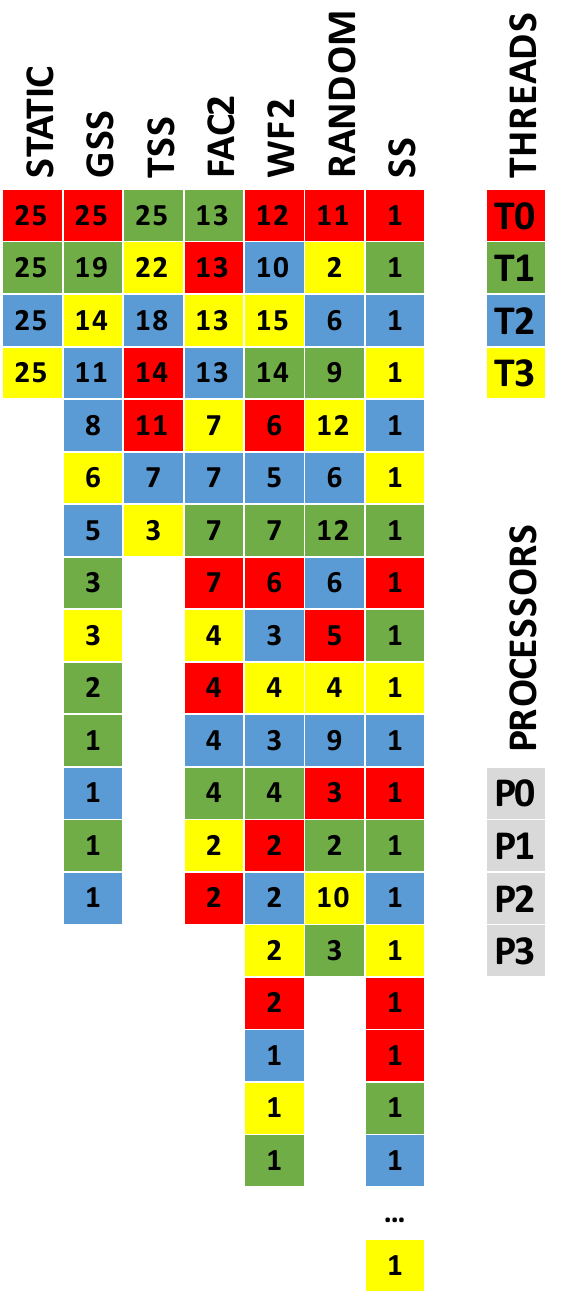}}%
\hfill
\subcaptionbox{Chunk execution\label{fig:DLS-execution}}
{\includegraphics[width=0.73\textwidth]{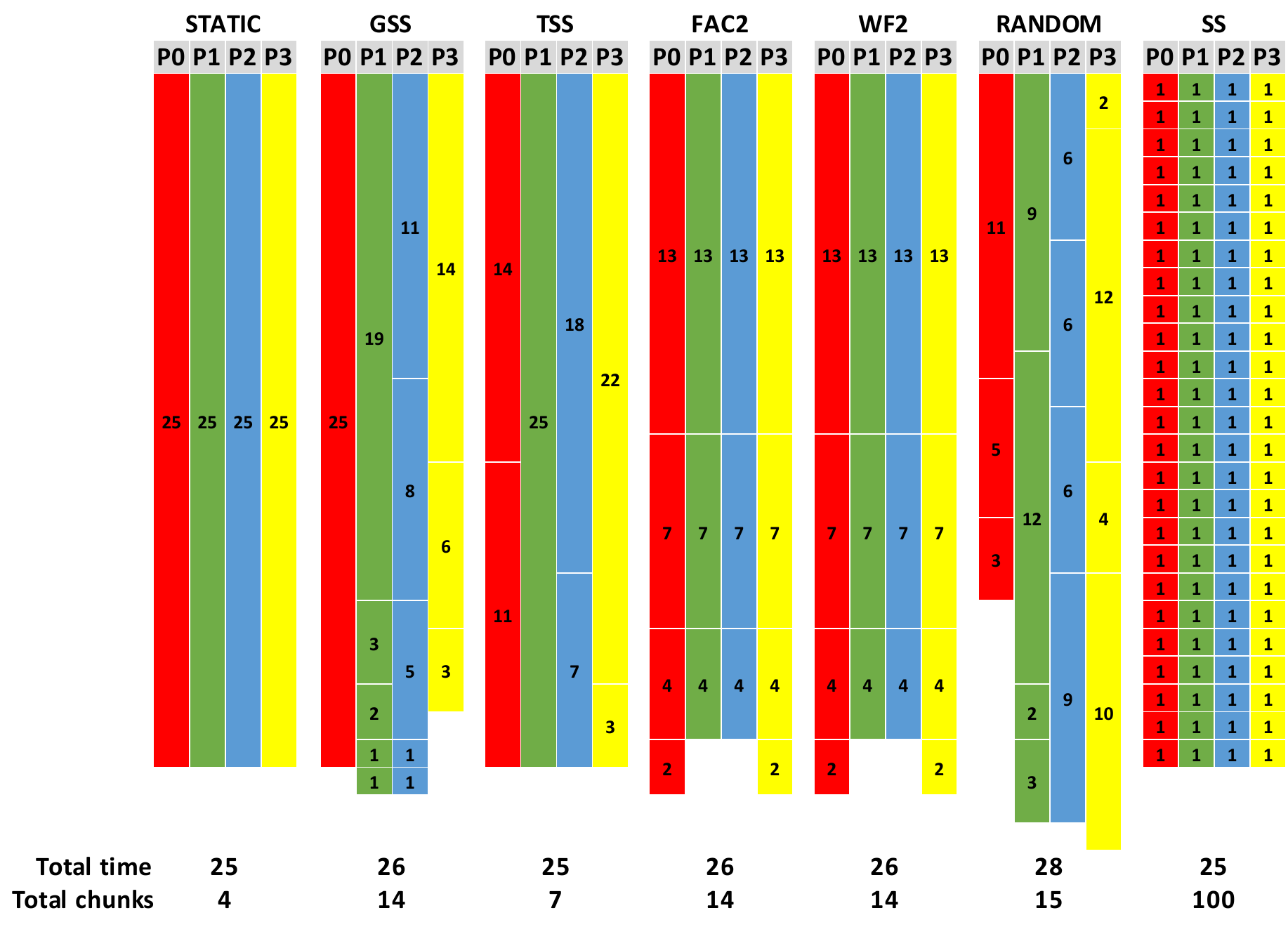}}%
\caption{The use of different DLS techniques for 
(a)~calculation of chunk sizes (numbers within colored rectangles) for 100~tasks and 
(b)~their execution over time using 4~threads~(T0,..,T3) assigned to 4~processing units~(P0,..,P3). 
While certain DLS achieve comparable ``absolute'' execution times, the incurred scheduling overhead, directly proportional with the number of self-scheduled chunks, may prohibitively degrade performance.}
\label{fig:dls-chunks-exec}
\end{figure}

A comparison of the prior existing and newly added OpenMP loop schedules is illustrated in Fig.~\ref{fig:dls-chunks-exec} for scheduling 100~(uniformly distributed) tasks on 4 homogeneous processors. 
\static denotes \texttt{schedule(static,chunk)}, \gss denotes \texttt{schedule(guided)}, and \selfsched denotes \texttt{schedule(dynamic,1)}.

The ordering of work requests made by the threads as illustrated in Fig.~\ref{fig:DLS-chunks} is only an instantiation of a dynamic process, and may change with every experiment repetition. 
Even though the scheduling overheads and allocation delays are not accounted for in this example (Fig.~\ref{fig:dls-chunks-exec}), one can easily identify the differences between the chunk sizes, the total number of chunks, their assignment order, and the impact of all these factors on the total execution time. 
STATIC and SS represent the two extremes of the \emph{load balance vs. scheduling overhead} trade-off (see Fig.~\ref{fig:ss-extremes}). 

The remaining DLS techniques are ordered by their efficiency in minimizing scheduling overhead and load imbalance.
The \texttt{schedule(guided,chunk)}\footnote{Here \texttt{chunk} denotes the smallest chunk size to be scheduled.} and \texttt{schedule(dynamic,chunk)}\footnote{Here \texttt{chunk} denotes a fixed chunk size to be scheduled.} versions of the respective \openmp schedules are not considered here due to the fact that they are simple variations of GSS and SS and fall between the two extremes of the trade-off illustrated in Fig.~\ref{fig:ss-extremes}, which, for reasons of clarity, only illustrates the schedules considered in this work.
\section{Evaluation of Selected DLS Techniques}\label{sec:dls-in-omp}
In its current state, the \openmp standard contains the two extremes of the scheduling spectrum and one self-scheduling example (see Fig.~\ref{fig:ss-extremes}).
While the state of the art has produced important additional scheduling methods over the last decades, the \openmp standard has not yet adopted these advances.

To evaluate the benefit of more advanced loop schedules, four additional schedules were added to an \openmp runtime library for the GNU compiler called \texttt{LaPeSD-libGOMP}\,\cite{LaPeSD-LibGOMP} and tested on molecular dynamics (MD) benchmarks, which are known use non-uniformly distributed \texttt{parallel for} loops resulting in imbalanced workloads (problem and algorithmic imbalance)\,\cite{FRAC:1995}.
CPU binding and pinning was used to inject additional load imbalance (system imbalance) in those applications.
The four added loop scheduling techniques were compared against the schedules available in current \openmp using this libGOMP version, the MD benchmarks and special measurement setup.

\begin{figure}[!h]
\centering
\includegraphics[width=0.95\textwidth]{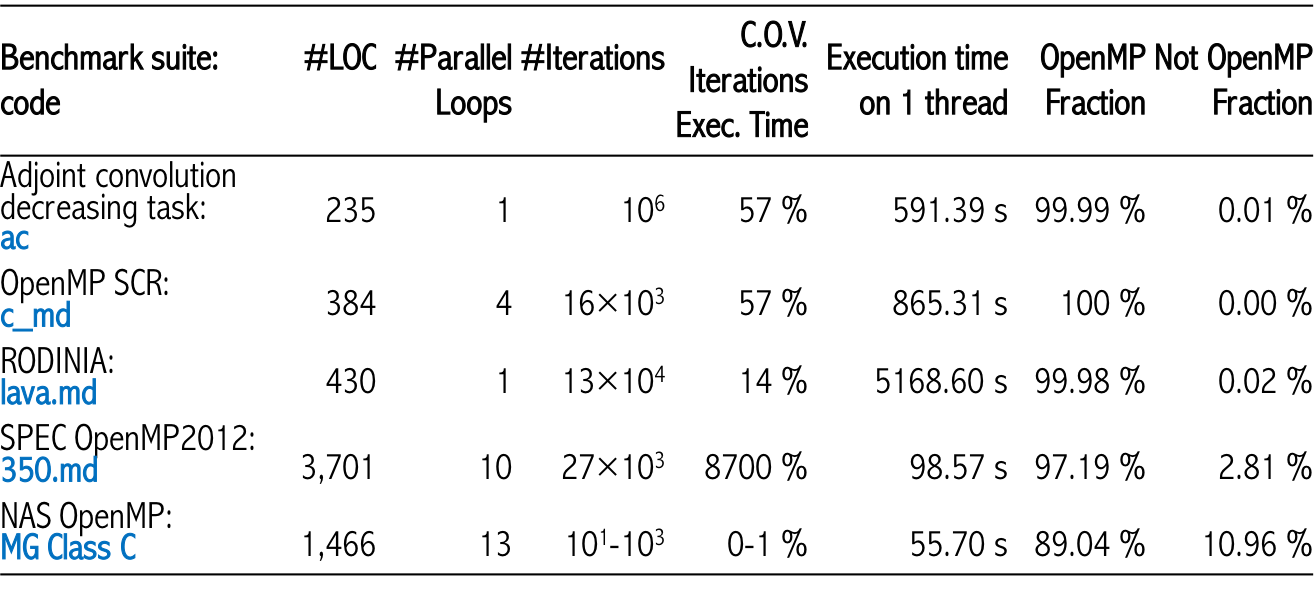}
\caption{\label{fig:benchmarks-doe}Characteristics of the selected OpenMP benchmarks. \#LOC denotes lines of code while C.O.V. denotes coefficient of variation.}
\end{figure}

\oursubsection{Benchmarks and schedule selection}
The set of selected OpenMP benchmarks is described in Fig.~\ref{fig:benchmarks-doe}.
This work considers applications simulating MD, which typically consist of large loops over complex molecular structures with variable degree of interactivity between different molecules.
While full MD applications are highly complex and challenging to be used for comparison in a performance testing environment, the comparatively short runtimes allowed an efficient exploration of the intended parameter space.
The choice of MD benchmarks is made from various well-established suites and includes: \cmd from the \openmp Source Code Repository (SCR) suite\,\cite{OMP-SCR:2005}, \lava from the RODINIA suite\,\cite{RODINIA:2010}, \nasmg (Class C) from the NAS OpenMP suite\,\cite{NAS:1991}, and \tfomd from the SPEC OpenMP 2012 suite\,\cite{SPEC-OMP:2012}.

An additional test application considered in this work is an in-house implementation of a linear algebra kernel, \emph{adjoint convolution} with decreasing task size, called \ac\footnote{In-house codes available at \url{https://bitbucket.org/PatrickABuder/libgomp/src}.}. 
This benchmark provides an ideal input to self-scheduling due to its high imbalance caused by the decreasing task size towards the end of the kernel.
Overall, the added benefit of using benchmarks is the avoidance of issues typically present in full production-ready codes, such as variations and influences of the MPI parallelization and I/O activities.

The four additional dynamic loop self-scheduling (DLS) techniques considered herein are \longtss (\tss)\,\cite{TSS:1993}, \longfacTWO~(\facTWO)\cite{FAC:1991}, \longwfTWO (\wfTWO)\,\cite{WF:1996}, and \longrandom (\random). 
While the literature offers more advanced DLS, this work considers DLS that require no additional runtime measurement during their execution, i.e.,  all used DLS are dynamic and \mbox{non-adaptive}\footnote{In non-adaptive dynamic scheduling, the schedule does not adapt to runtime performance observations other than to the dynamic consumption of tasks.}. Recall that the \facTWO and \wfTWO methods evolved from the probabilistic analysis behind FAC and WF, respectively.
Moreover, recall that \wfTWO can employ workload balancing information specified by the user, such as the capabilities of a heterogeneous hardware configuration.

\oursubsection{Prototype implementation}
The development of a dedicated \openmp compiler and runtime implementation is beyond the scope of this work.
Therefore, the GNU implementation of \openmp is used, as the scheduling mechanisms are independent from the actual compilation of \openmp constructs and of their separate implementation in a separate runtime library.
For the implementation\footnotemark[3] of the additional loop schedules the \texttt{LaPeSD-libGOMP} is used as basis.\hfill

\begin{figure}[!h]
\centering
\includegraphics[width=0.45\textwidth]{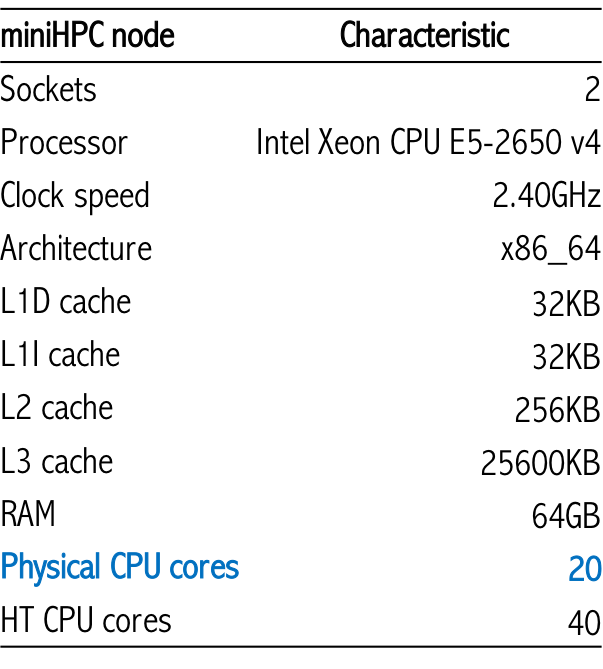}
\caption{\label{fig:minihpc}Hardware characteristics of a single node of the \emph{miniHPC} system used in the experiments.}
\end{figure}

\oursubsection{Measurement setup and evaluation policy}
The experiments for this work were conducted on a single node of the \emph{miniHPC} system at the University of Basel (see Fig.~\ref{fig:minihpc}), a fully-controlled research system with a dual-socket 10-core Intel Xeon E5-2640 v4, with both hyper-threading and TurboBoost enabled.
In all experiments the hyper-threaded cores with ID 20-39 were left idle.

All experiments were conducted exclusively, each experiment being repeated 20 times.
For the results shown in this work, the \emph{median} and the \emph{standard deviation} of the execution times were gathered for all repetitions.
As discussed in Section\,\secref{sec:revisiting}, load imbalance has three potential causes: 
either the workload of the iteration space is imbalanced, 
the work assignment is imbalanced, or 
the hardware is imbalanced, e.g., as in a heterogeneous system.
Since the work assignment (scheduling) is defined by the used schedule clause, any load imbalance originates in either or in both, variability in the benchmark itself or from performance variability in the hardware.

\begin{figure}[hp]
\centering
\centering{}
\begin{subfigure}[b]{0.48\textwidth}
\includegraphics[width=\textwidth]{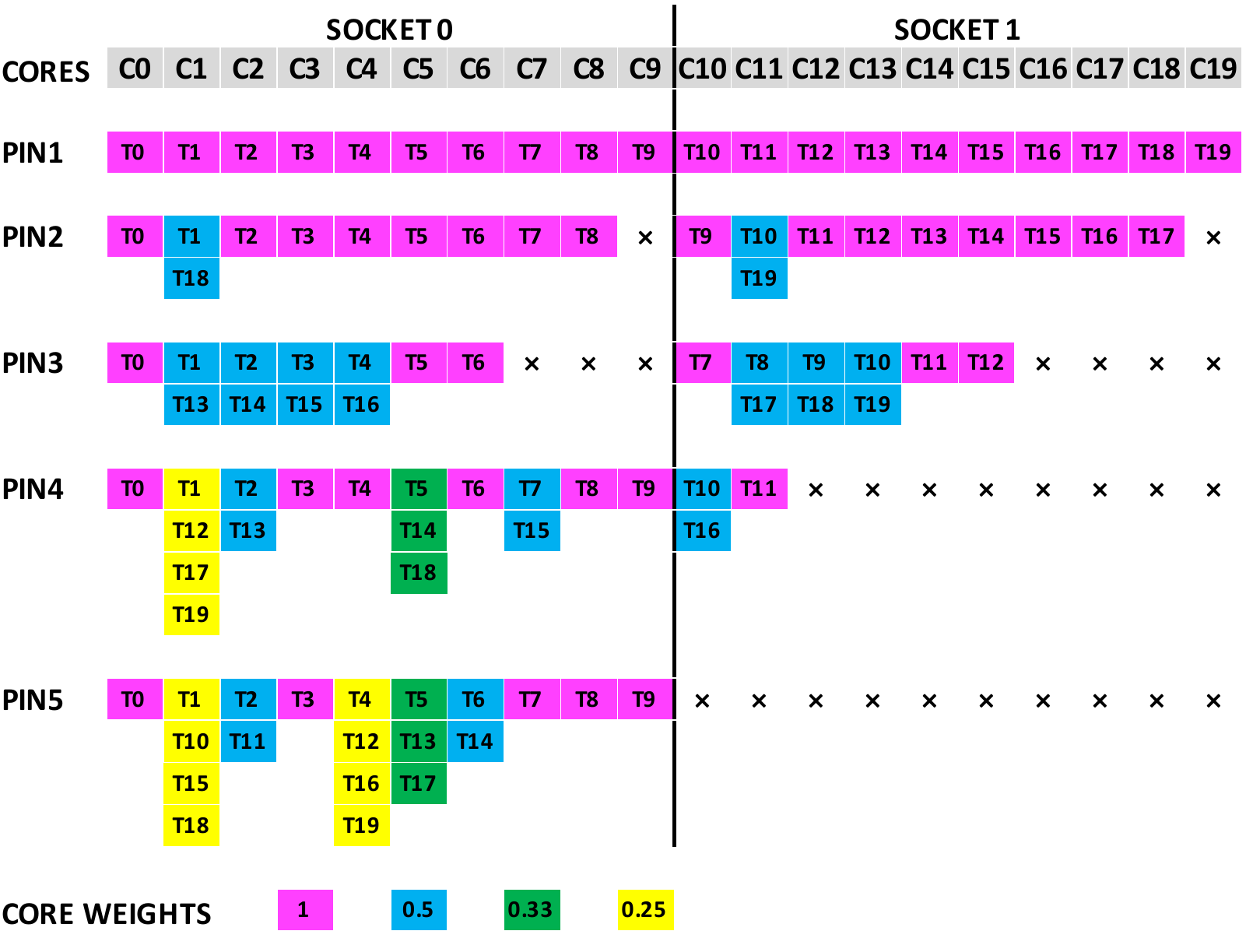}
\caption{\label{fig:pinnings} Pinning patterns}
\end{subfigure}\hfill%
\begin{subfigure}[b]{0.48\textwidth}
\includegraphics[width=\textwidth]{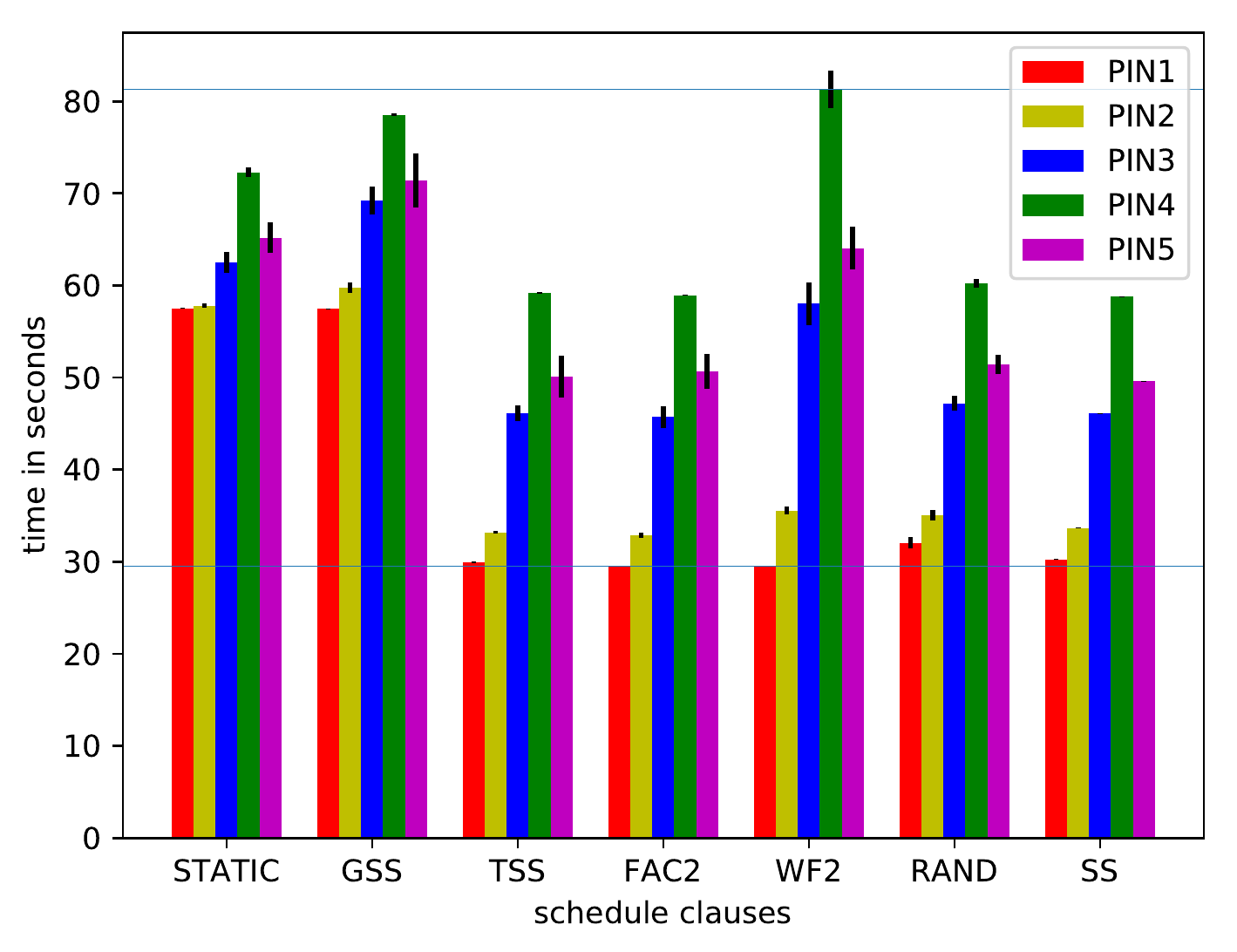}
\caption{\ac}
\label{fig:AC}
\end{subfigure}\\
\begin{subfigure}[b]{0.48\textwidth}
\includegraphics[width=\textwidth]{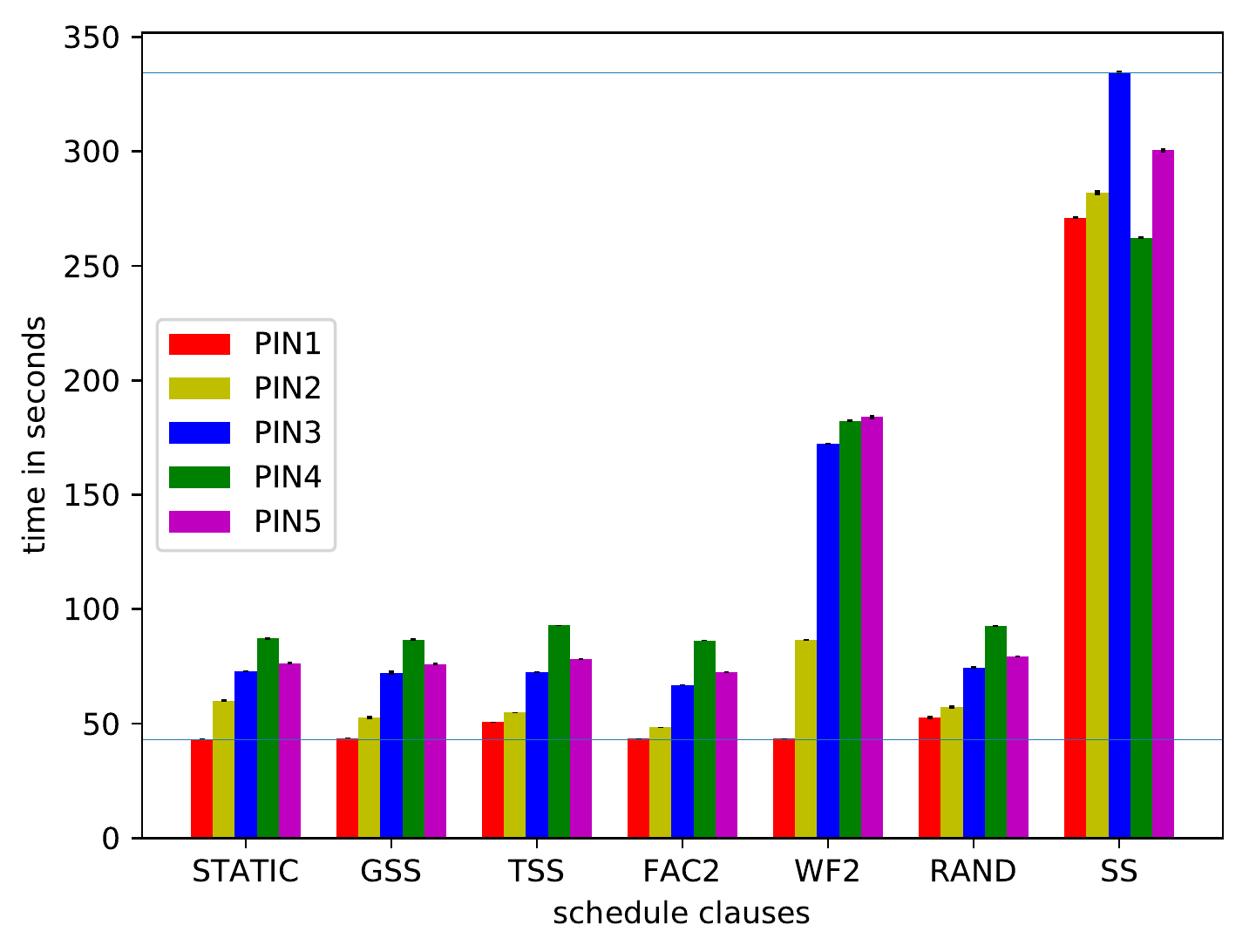}
\caption{\cmd}
\label{fig:c_md}
\end{subfigure}%
\hfill%
\begin{subfigure}[b]{0.48\textwidth}
\includegraphics[width=\textwidth]{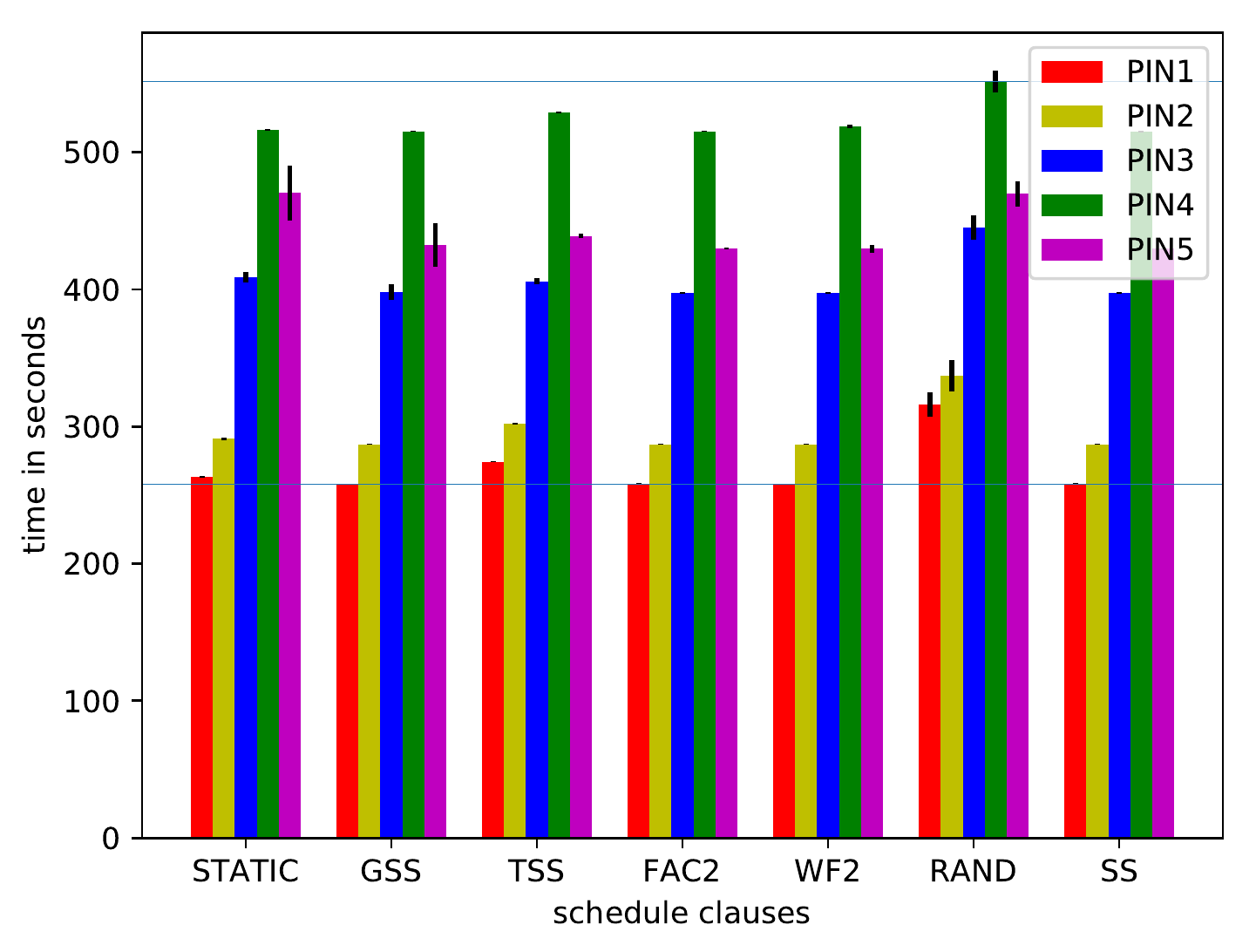}
\caption{\lava}
\label{fig:lavaMD}
\end{subfigure}
\\
\begin{subfigure}[b]{0.48\textwidth}
\includegraphics[width=\textwidth]{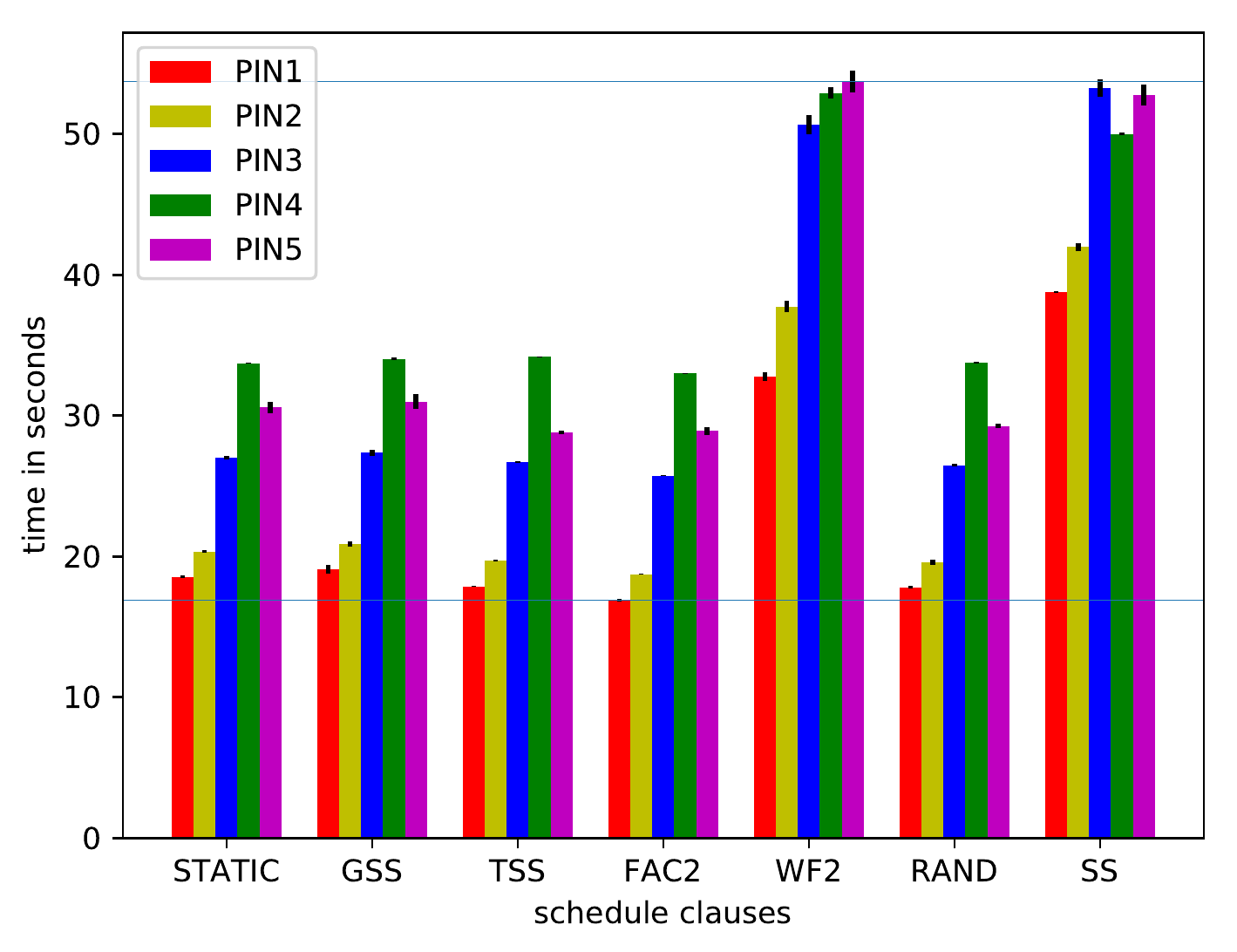}
\caption{\tfomd}
\label{fig:350md}
\end{subfigure}
\hfill
\begin{subfigure}[b]{0.48\textwidth}
\includegraphics[width=\textwidth]{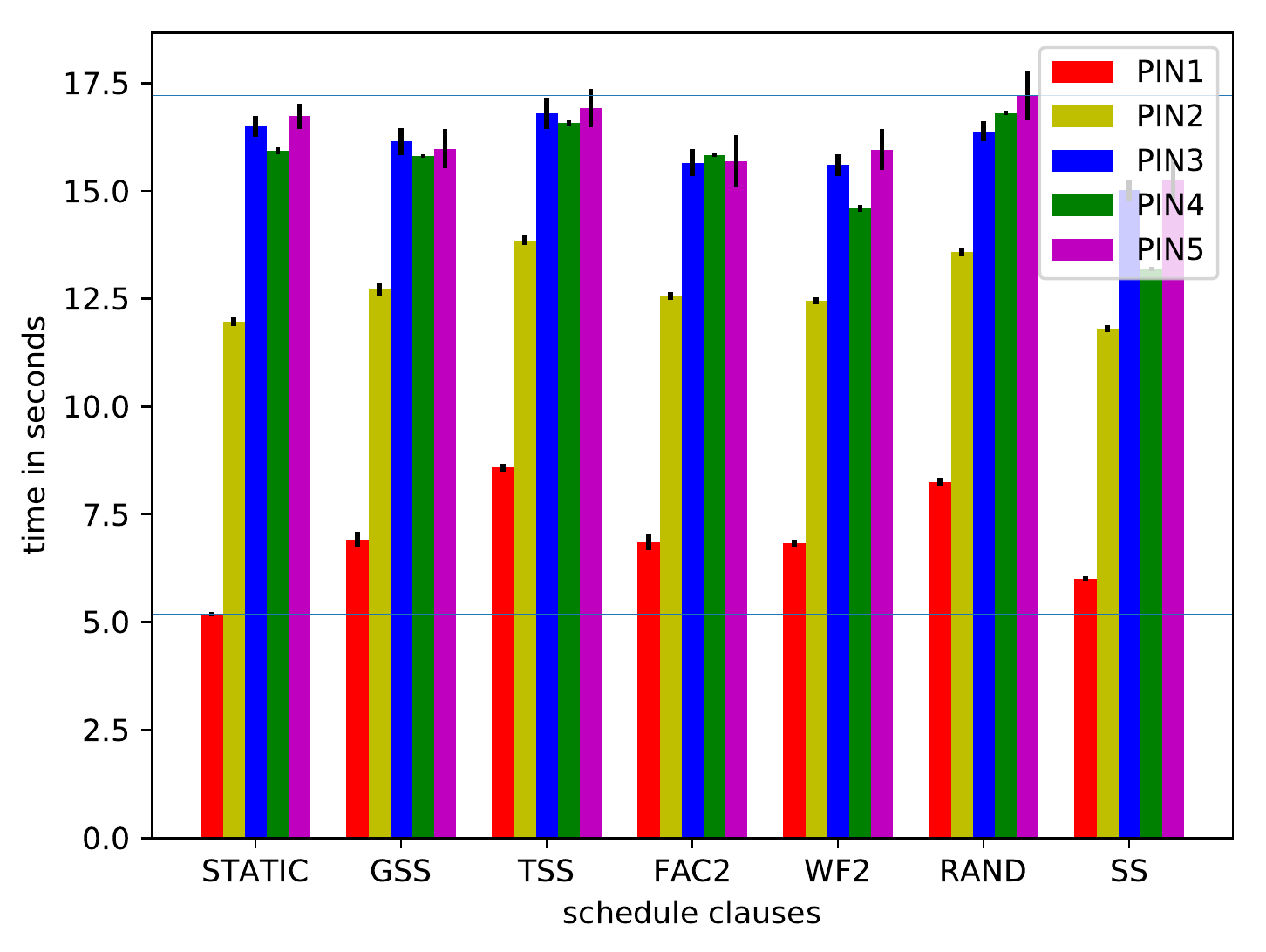}
\caption{\nasmg}
\label{fig:nas}
\end{subfigure}
\caption{Pinning Patterns and Measurements}
\label{fig:measurements}
\parbox{\linewidth}{\footnotesize{}\textbf{(a)} Pinning patterns: T0-T19 represent \openmp threads, C0-C19 represent physical cores. Core weights depict fraction of the performance of a single core available for a given thread. \textbf{(b)-(f)} The existing \openmp schedules, \static, \gss, and \selfsched, as well as the additional \tss, \facTWO, \wfTWO, and \random schedules from the present own implementation were evaluated for each benchmark. Minimum and maximum runtimes are marked by horizontal blue lines.}
\end{figure}

For this work, the principal load imbalance originates in the workloads in \openmp codes of the MD benchmarks.
However, as the benchmarks, in their original form, are rather limited in loop sizes and input data, the inherent imbalance is not extreme.
To highlight the impact of the loop scheduling choice, additional artificial hardware irregularity is added by omitting specific CPU cores for use in the experiments and by binding the 20 OpenMP threads to the remaining cores, reducing the achieved performance of those threads in the process. 
\Figref{fig:pinnings} provides an overview of the five pinning and oversubscription configurations used.
In the experiments, \emph{PIN1} to \emph{PIN5} were used as the pinning and binding schemes.
\emph{{PIN1}} uses all the cores of the system and binds each thread to a single  core.
\emph{{PIN2}} considers cores 9 and 18 as not present in the system and binds two additional threads to cores 1 and 11 (respectively), reducing the performance of threads 1,10,18, and 19.
\emph{PIN3} increases the hardware load imbalance by removing cores 7,8,16,17 and 18 from the set of available CPU cores, hence, reducing the performance of threads 1,2,3,4,8,9,10,13,14,15,16, and~18.
\emph{PIN4} adds further inhomogeneity into the system; cores 12 to 19 remain idle, while the twenty threads are bound to cores 1,2,5,7, and 10. 
This creates four classes of overloaded CPUs: with 4 threads, with 3 threads, and with 2 threads. 
The last setup, \emph{PIN5}, uses only one (out of two) socket.
Note that pinnings \emph{PIN3}-\emph{PIN5} explicitly avoid a symmetric configuration to increase the level of imbalance in the system.
This decreases the efficiency of threads being executed on those cores, hence, implicitly generating a substantial compound load imbalance for the application.
An efficient load balancing scheme is expected to mitigate such performance variations due to compound load imbalance such as results from the \emph{PIN3}-\emph{PIN5} setups.
\oursubsection{Experimental evaluation}
The results of all selected benchmarks are shown in \figref{fig:AC} to \ref{fig:nas}. 
For this work, a specific schedule is considered to be advantageous if it provides a performance benefit directly to the parallel loops or if it provides a performance benefit in coping with the hardware heterogeneity.
The \ac has approximately 10\mbox{\sc{e}}6 iterations to schedule, having a coefficient of variation of the iterations execution times (CV) of 57\% and is, as such, a suitable example for \mbox{self-scheduling}.
As expected, the \static and \gss schedules struggle to cope with the highly irregular workload, other schedules with a less strict scheduling regiment perform well, with \facTWO performing the best. 
The \static exhibits the least sensitivity to the different \emph{PIN}-regiments with \wfTWO exhibiting the highest sensitivity.
For \ac, \facTWO offers the overall best performance in all configurations with \selfsched offering comparable performance for \emph{PIN4} and \emph{PIN5} as it can offset its high overhead by achieving the best load balancing effect.

For \cmd with its roughly 16\mbox{\sc{e}}3 iterations and CV of 57\%, the \facTWO offers the best performance in the regular setup (\emph{PIN1}) with \static, \gss and \wfTWO coming very close; \tss and \random provide a slightly lower performance while the \selfsched exhibits a 5 times lower performance -- likely due to the small iteration space and high scheduling overhead.
For the imbalanced PIN configurations all schedules, except \wfTWO and \selfsched, fare well, offering comparable performance.
\wfTWO is substantially affected by the hardware-induced load imbalance -- likely due to its initial calibration with the attempt to account for the hardware imbalance and, subsequently, to the lack of iterations to benefit form the high setup-overhead.
The \selfsched exhibits an interesting behavior for this benchmark: while it suffers un-proportionally for \emph{PIN3}, it performs better for \emph{PIN4}, even against its \emph{PIN1} configuration.

The \lava benchmark has a comparable low peak CV of 14\% in its 13\mbox{\sc{e}}4 iterations.
For this reason, it is not surprising that all but the \random schedule offer comparable performance; \random even induces a load imbalance into the benchmark.
For the \emph{PIN2}-\emph{PIN5} configurations there is little difference between the schedules, again with the exception of \random.

For the \tfomd the iteration count is about 27\mbox{\sc{e}}3 with a high CV of 8700\%.
Again \facTWO has the best performance advantage, making use of the high variations in the loop iterations.
The \static, \gss, \tss, and \random schedules also perform well and reach about 10\% of the peak performance.
\selfsched and \wfTWO provide the lowest performance, failing to exploit the high CV in the workload due to the overhead of scheduling a much larger number of chunks. 
 
The \nasmg benchmark has a very low loop iteration space, between 2 and 1k, with a very strongly variable CV between 0, i.e., no variation, and 1.
However, many such loops are instantiated in this benchmark which results in approximately 5k instantiations. 
Here the \static schedule offers the best performance for the \emph{PIN1} configuration as its overhead is minimal and the iteration space and, thus, the aggregated load imbalance is low.
For this reason, the \selfsched offers the best performance in the hardware imbalanced \emph{PIN2}-\emph{PIN5} configurations, as it has few iterations to schedule and a high hardware-based imbalance to mitigate.
For the same reason, the \wfTWO also performs much better, even though the cost to calculate a chunk is higher -- it is amortized by the well balanced scheduling.

\oursubsection{Discussion}
In this work an outline of a prototype implementation of additional schedules in \openmp has been provided together with an evaluation thereof using various benchmarks and hardware setups.
While it is clear that the additional schedules do not provide a one-fits-all solution, the greater variety of choice offers an opportunity for improvement in many instances.
The precise circumstances of such a benefit must be taken into account when deciding which scheduling choice is the best.
The number of iterations to be scheduled, the chunk sizes, and the overheads associated with each approach factor heavily into the decision of which schedule to use.
The experience from the prototype implementation 
and the subsequent test on benchmarks, support the original hypothesis that indeed, \emph{additional schedules provide benefit.}
While a single schedule was applied for all work-sharing loops in a code, a more diverse use of the scheduling clause will further increase the benefit.
This can, for example, be achieved by leveraging domain expert knowledge, not considered in this work.
A critical observation is that many loops from the considered benchmarks \emph{did not use any scheduling clause} in the various work-sharing loops.
It is well known that explicitly using the \openmp \texttt{schedule} is critical for improving performance. 
This can bee seen in \cmd, \ac, \tfomd and \nasmg, where the performance difference between the fastest and slowest schedule exceeded a factor of two.
\section{Summary}\label{sec:conclusion}

This work revisits the scheduling of \openmp work-sharing loops and offers an overview of the current state-of-the-art self-scheduling techniques.
Scheduling is a performance critical aspect of loops that is currently receiving insufficient attention.
This work investigates alternative state-of-the-art loop self-scheduling schemes from the literature and reflects on their use in \openmp loop scheduling.
In Section\,\secref{sec:revisiting}, it was shown that the existing \openmp schemes \static and \selfsched are only the extremes of a broad spectrum of state-of-the-art loop self-scheduling methods, and that \gss represents a well-known, yet obsolete variant within that spectrum, with many newer more efficient schedules being available.
Using a selection of more recent loop-scheduling techniques, prototype implementations were developed for the GNU \openmp runtime library and used in viability experiments.
Section~\secref{sec:dls-in-omp} provides an overview of this effort as well as measurements results.
These results show that more recent loop self-scheduling techniques exhibit, in three out of five test cases, an improved performance than the schedules already available in \openmp.
With the theoretical underpinnings of loop self-scheduling and the results presented, it is clear that \openmp should include a greater variety of loop schedules.

Fueled by the increasing heterogeneity of hardware, i.e., classic CPUs in combination with accelerators and future heterogeneous CPUs, a fixed set of schedules will not suffice, even if it were to outperform all currently available scheduling techniques (in the absence of load imbalance).
Therefore, the \openmp community is invited to address this issue, especially with the addition of tasking that brings additional challenges in the area of scheduling in \openmp.
From this perspective, an interface for user-defined schedules, comparable to\,\cite{UDefSched:2017}, would be preferable, as it allows users and library developers to add scheduling techniques without having to update the \openmp standard every time a new approach is developed.
In addition, the interface should enable developers to provide the runtime system with more expert knowledge and information regarding the variability of the iteration space of the loops in their applications.

Future work will explore more schedules with more complex underlying models, such as the adaptive self-schedules, which employ feedback loops and require performance measurements. 
The applicability of self-scheduling is also of high interest, as for example the global load balancing of task scheduling in task-loops remains unaddressed.
Further research in the domain of heterogeneity and NUMA and its impact on scheduling is also planned. 

\section*{Acknowledgment}
This work is partly funded by the Swiss National Science Foundation in the context of the ``Multi-level Scheduling in Large Scale High Performance Computers'' (MLS) grant, number 169123, and by the Hessian State Ministry of Higher Education by granting the ``Hessian Competence Center for High Performance Computing''.
\vspace{-.25cm}
%
%
%
%
\bibliographystyle{abbrv}
\bibliography{loop_scheduling}

\end{document}